\documentclass[12pt,twoside]{article} 
\usepackage{epsfig} 
\newcommand{\figg}[5]{\begin{figure}[t]
  \centerline{\epsfig{file=#1.ps,height=#2cm,width=#3cm,silent=}}
        \caption{#4}
        \label{#5}
        \end{figure}}
\title{Relativistic contraction of an accelerated rod}
\author{Hrvoje Nikoli\'c  \\
Theoretical Physics Division, Rudjer Bo\v{s}kovi\'{c} Institute, \\
P.O.B. 1016, HR-10001 Zagreb, Croatia \\
{\normalsize hrvoje@faust.irb.hr} \\
\makebox[1in]{} \\
}
\date{\today}
\begin{document}
\maketitle
\begin{abstract}
The relativistic motion of an arbitrary point of an accelerated 
rigid rod is discussed 
for the case when velocity and acceleration are directed along the
rod's length. The discussion includes the case of a time-dependent
force applied to a single point, as well as a time-independent  
force arbitrary distributed along the rod. 
 The time dependence
 of the rod's relativistic length depends on the application point 
 of the force, but after the termination of acceleration 
the final velocity and length 
 do not depend on it. An observer on a uniformly accelerated rod
 feels an inertial force which decreases in the direction of acceleration.
The influence of non-rigidity of realistic rods on our 
results is qualitatively discussed.  
\end{abstract}

%
 
\section{Introduction}

There are several articles which discuss relativistic properties of 
accelerated rods for the case when the force is time-independent and applied 
to a single point on the rod. Cavalleri and Spinelli [1] found two 
important results for such a case. First, the application point 
accelerates in the same way as it would accelerate if all mass
of the rod were concentrated in this point. Second, ``a rod, pushed by a 
force, accelerates less than the same rod pulled by the same force". 
Some similar results were found by Nordtvedt [2] and 
Gr\rlap/on [3], who concluded that ``a rocket ship 
can be accelerated to higher speeds with a given engine by putting the 
engine in the nose of the rocket". We agree with the first statement 
in quotation marks, but we disagree with the second one. At first 
sight, the second statement in quotation marks may seem to be a consequence 
of the first one. On the other hand, the second statement cannot be  
consistent with the conservation of energy. We resolve the paradox by 
generalizing the analysis to time-dependent forces. As an example 
we consider the case of 
a uniformly accelerated rod during a finite time interval, after which the
force turns off. It appears that although the motion of the rod 
depends on the application point of the force, the final velocity and 
relativistic length after the termination of acceleration do not depend 
on it. From the point of view of an inertial observer at rest, a pushed 
rod accelerates slower than a pulled one, but the acceleration of a pushed 
rod lasts longer than that of a pulled one. 

In this article we also generalize the first result of [1] by 
considering the case of many time-independent forces applied to various 
points. There is one point which accelerates
in the same way as it would accelerate if all forces were applied to
this point. We find that this point is given by (\ref{g8}). 

The paper is organized as follows: In Section 2 we present a general method 
of determining the motion of an arbitrary point of a non-rotating 
rigid rod, when the motion of 
one of its points is known. We also present a general method of determining the
motion of a point on the rigid rod, to which a single time-dependent force is
applied. In Section 3 we apply this general formalism to the case of 
a uniformly accelerated rod during a finite time interval, after which the
force turns off. In Section 4 we discuss the physical meaning of the results 
obtained in Section 3. In Section 5 we analyze the case of many
time-independent forces applied to various points. 
In Section 6 we give a qualitative 
discussion of how the non-rigidity of realistic rods alters our 
analysis and find conditions under which our analysis is 
still valid, at least approximately.
Section 7 is devoted to concluding remarks.  
    
\section{The length of a rigid accelerated rod}

Let us consider a rod whose velocity and acceleration are directed along its 
length, but are otherwise arbitrary functions of time. 
We assume that the accelerated rod is rigid, which means that an
observer located on the rod does not observe any change of the rod's length. 
(In Section 6 we discuss the validity of such an assumption.) Since the rod is
rigid and does not rotate, it is enough to know how one particular point 
of the rod (labeled for example by $A$) changes its position with time.      
Let $S$ be a stationary inertial frame and $S'$ the accelerated frame of an 
observer on the rod. We assume that we know the function $x_A(t_A)$, so 
we also know the velocity
\begin{equation}\label{8}
 v(t_A)=\frac{d x_A(t_A)}{d t_A} \; .
\end{equation}
Later we consider how the function $x_A(t_A)$ can be found if it is known how
the forces $F(t')$ act on the rod. The function (\ref{8}) defines the 
infinite sequence of 
comoving inertial frames $S'(t_A)$. The rod is instantaneously at rest for 
an observer in this frame. This means that he observes no contraction, i.e., 
$L_0 =x_B' -x_A'$, where $L_0$ is the proper length of the rod, while $A$ and $B$ 
label the back and front ends of the rod, respectively. He observes both ends 
at the same instant, so $t_B' -t_A' =0$. From the Lorentz transformations 
\begin{equation}\label{1}
 x_{A,B}=\gamma_v (t_A) (x_{A,B}' +v(t_A) t_{A,B}') \; , \;\;\;\;\;
 t_{A,B}=\gamma_v (t_A) \left(t_{A,B}' +\displaystyle\frac{v(t_A)}{c^2} x_{A,B}'
   \right) \; ,  
\end{equation}
where $\gamma_v (t_A)=(1-v^2(t_A)/c^2)^{-1/2}$, we obtain 
\begin{equation}\label{9.1}
 x_B -x_A =L_0 \gamma_v(t_A) \; ,
\end{equation}
\begin{equation}\label{9.2}
 t_B -t_A =L_0 \gamma_v(t_A) \frac{v(t_A)}{c^2} \; .
\end{equation}
From (\ref{9.1}) and the known functions $x_A(t_A)$ and (\ref{8}) we can find
the function $x_B(t_A)$. From (\ref{9.2}) and the known function (\ref{8}) 
we can find the function $t_A(t_B)$. Thus we find the function 
\begin{equation}\label{10}
 x_B(t_A(t_B)) \equiv \tilde{x}_B (t_B) \; .
\end{equation}
To determine how the rod's length changes with time for an observer in 
$S$, both ends of the rod must be observed at the same instant, so 
$t_B =t_A\equiv t$. 
Thus the length as a function of time is given by 
\begin{equation}\label{11}
 L(t)=\tilde{x}_B (t)-x_A (t) \; .
\end{equation}

Let us now see how velocity (\ref{8}) can be found if the force
$F(t_A')$ applied to the point $A$ is known. $F$ is the force as seen by an
observer in $S'$. We introduce the quantity 
\begin{equation}\label{t1}
 a(t_A')=F(t_A')/m(t_A') \; ,
\end{equation}
which we call acceleration, having in mind that this would be the 
second time derivative of a position only in the nonrelativistic limit. Here 
$m(t_A')$ is the proper mass of the rod, which, in general, can also change with 
time, for example by loosing the fuel of a rocket engine. As shown  
in [1], if there is only one force, applied to  
a specific point on an elastic body, and if $F$ and $m$ do not vary with time, 
then this point moves in the same way as it would move if all mass
of the body were concentrated in this point. If acceleration changes with
time slowly enough, then this is approximately true for a time-dependent 
acceleration as well. Later we discuss the conditions for validity of 
such an approximation. 
Here we assume that these conditions are fulfilled. The application point 
is labeled by $A$. Thus, by a straightforward application of the velocity 
addition formula, we find that  
the infinitesimal change of velocity is given by 
\begin{equation}\label{t2}
 u(t_A' +d t_A')=\frac{u(t_A') +a(t_A') d t_A'}{1+\displaystyle\frac{ 
   u(t_A')a(t_A')d t_A'}{c^2}}=
 u(t_A') +\left( 1-\frac{u^2(t_A')}{c^2}\right) a(t_A') d t_A' \; ,
\end{equation}
where $u(t_A')$ is velocity defined in such a way that
$u(t_A'(t_A))=v(t_A)$. 
Since $u(t_A' +d t_A')=u(t_A') +du$, this leads to the differential equation 
\begin{equation}\label{t3}
 \frac{d u(t_A')}{d t_A'} = \left( 1-\frac{u^2(t_A')}{c^2}\right) a(t_A') \; , 
\end{equation}
which can be easily integrated, since $a(t_A')$ is the known function by
assumption. Thus we find the function $u(t_A')$. To find the function
$v(t_A)$, we must find the function $t_A'(t_A)$. We find this from the 
infinitesimal Lorentz transformation
\begin{equation}\label{t4}
 dt_A =\frac{dt_A' +\displaystyle\frac{u(t_A')}{c^2} dx_A'}
       {\sqrt{1-u^2(t_A')/c^2}} \; .
\end{equation}
The point on the rod labeled by $A$ does not change, i.e., $dx_A' =0$, so 
(\ref{t4}) can be integrated as 
\begin{equation}\label{t5}
 t_A =\int \frac{dt_A'}{\sqrt{1-u^2(t_A')/c^2}} \; ,
\end{equation}
which gives a function $t_A =f(t_A')$ and thus $t_A'=f^{-1}(t_A)$.

It is also interesting to see how the length of an unaccelerated
rod changes with
time from the point of view of an accelerated observer. The generalized Lorentz 
transformations between an inertial frame and an accelerated frame, as 
shown by Nelson [4], are given by 
\begin{equation}\label{n12}
 x=\gamma_u(t')x' +\int_{0}^{t'} \gamma_u(t')u(t') dt' \; , \;\;\;\;\;
 t=\frac{\gamma_u(t')u(t')}{c^2}x' +\int_{0}^{t'} \gamma_u(t')dt' \; ,
\end{equation}
where $\gamma_u(t')=(1-u^2(t')/c^2)^{-1/2}$. Now the proper length of the rod 
is $L_0 =x_2 -x_1$. The accelerated observer observes both ends
at the same instant, so $t_2' =t_1'$. The length that he observes is 
$L'=x_2' -x_1'$, so from (\ref{n12}) we find
\begin{equation}\label{n3}
 L'(t')=\frac{L_0}{\gamma_u(t')} \; .
\end{equation}  

\section{Uniformly accelerated rod during a finite time interval}

In the preceding section we have made a very general analysis. Here we want to
illustrate these results on a simple realistic example, in order to
understand the physical meaning of these general results. We consider the
case of a rod which is at rest for $t<0$, but at $t=0$ it turns on its engine 
which gives the constant acceleration $a$ to the application point 
during a finite time interval $T'$, after which the engine turns off. 
From (\ref{t3}) and (\ref{t5}) for $t_A' <T'$ we find
\begin{equation}\label{tt1}
 u(t_A')=c\: {\rm tgh}\frac{at_A'}{c} \; ,
\end{equation}
\begin{equation}\label{tt3}
 t_A(t_A')=\frac{c}{a}\: {\rm sh}\frac{at_A'}{c} \; , 
\end{equation}  
and thus
\begin{equation}\label{12}
 v_A(t_A)=\left\{
 \begin{array}{l}
  \displaystyle\frac{at_A}{\sqrt{1+(at_A/c)^2}} 
   \; , \;\;\;\; 0\leq t_A\leq T \; , \\
     \\ 
  \displaystyle\frac{aT}{\sqrt{1+(aT/c)^2}} \; , \;\;\;\; t_A\geq T \; , 
 \end{array}\right.
\end{equation}
where 
\begin{equation}\label{12.1}
 T=\frac{c}{a}\: {\rm sh}\frac{aT'}{c} \; . 
\end{equation}
With the initial condition $x_A(t_A=0)=0$ we obtain
\begin{equation}\label{13}
 x_A(t_A)=\left\{
 \begin{array}{l}
  \sqrt{(c^2/a)^2 +(ct_A)^2} -c^2/a \; , \;\;\;\; 0\leq t_A\leq T  \; , \\
    \\
  \displaystyle\frac{aTt_A}{\sqrt{1+(aT/c)^2}} +\displaystyle\frac{c^2}{a}\left( 
    \displaystyle\frac{1}{\sqrt{1+(aT/c)^2}} -1\right) \; , \;\;\;\; t_A\geq T \; . 
 \end{array}\right.
\end{equation}
The rest of job is described by the procedure given from (\ref{9.1}) to
(\ref{11}). Thus we find 
\begin{equation}\label{14}
 t_A(t_B)=\left\{
 \begin{array}{l}
  \displaystyle\frac{t_B}{1+aL_0/c^2} 
   \; , \;\;\;\; 0\leq t_B \leq T_+ \; , \\
        \\ 
  t_B - aL_0T/c^2 \; , \;\;\;\; t_B \geq T_+ \; ,  
 \end{array}\right.
\end{equation}
\vspace{0.5cm}
\begin{equation}\label{15}
 \tilde{x}_B(t_B)=\left\{
 \begin{array}{l}
  \sqrt{1+\displaystyle\frac{(at_B/c)^2}{(1+aL_0/c^2)^2}}\left(
  \displaystyle\frac{c^2}{a} 
   +L_0\right) 
   -\displaystyle\frac{c^2}{a} \; , \;\;\;\; 0\leq t_B \leq T_+ \; , \\
     \\
  \displaystyle\frac{1}{\sqrt{1+(aT/c)^2}}
   \left( \displaystyle\frac{c^2}{a} +L_0 +aTt_B \right) 
   -\displaystyle\frac{c^2}{a} \; , \;\; t_B \geq T_+ \; ,
 \end{array}\right.
\end{equation}
\vspace{0.5cm}
\begin{equation}\label{17}
 L(t)=\left\{
 \begin{array}{l}
  \sqrt{1+\displaystyle\frac{(at/c)^2}{(1+aL_0/c^2)^2}}
   \left( \displaystyle\frac{c^2}{a} +L_0\right) 
   -\displaystyle\frac{c^2}{a}\sqrt{1+(at/c)^2} \; , \;\;\;\; 0\leq t \leq T \; , \\
      \\
  \sqrt{1+\displaystyle\frac{(at/c)^2}{(1+aL_0/c^2)^2}}
   \left( \displaystyle\frac{c^2}{a} +L_0\right)
   -\displaystyle\frac{1}{\sqrt{1+(aT/c)^2}}
   \left( \displaystyle\frac{c^2}{a} +aTt \right) \; , \\ 
   \;\;\;\;\;\;\;\;\;\;\;\;\;\;\;\;\;\;\;\;\;\;\;\;\;\;\;\;\;\;\;\;\;\;\;
  \;\;\;\;\;\;\;\;\;\;\;\;\;\;\;\;\;\;\;\;\;\;\;\;\;\;\;\;\;\;\;\;\;\;\;\;
   T\leq t \leq T_+ \; , \\
  L_f \; , \;\;\;\; t\geq T_+ \; ,
 \end{array}\right.
\end{equation}
where $T_{\pm}=T(1\pm aL_0/c^2)$, while $L_f =L_0/ \sqrt{1+(aT/c)^2}$ is the 
final length. 
Note that (\ref{17}) differs from the result which one could expect from the naive 
generalization of the Lorentz-Fitzgerald formula
\begin{equation}\label{18}
 L(t)=L_0\sqrt{1-v^2(t)/c^2}=\frac{L_0}{\sqrt{1+(at/c)^2}} \; , \;\;\;\; 
 0\leq t \leq T \; .
\end{equation}
  
Formula (\ref{17}) was obtained for the case when the force is applied to the 
back end of the rod. In other words, this is the result for a pushed rod. 
The analysis for a pulled rod is similar and the result is 
\begin{equation}\label{17.1}
 L(t)=\left\{
 \begin{array}{l}
  \displaystyle\frac{c^2}{a}\sqrt{1+(at/c)^2} -
   \sqrt{1+\displaystyle\frac{(at/c)^2}{(1-aL_0/c^2)^2}}
   \left( \displaystyle\frac{c^2}{a} -L_0\right)
   \; , \;\;\;\; 0\leq t \leq T_- \; , \\
         \\
  \displaystyle\frac{c^2}{a}\sqrt{1+(at/c)^2} 
   -\displaystyle\frac{1}{\sqrt{1+(aT/c)^2}}\left(
   \frac{c^2}{a} -L_0 +aTt \right) \; , \;\;\;\; T_- \leq t\leq T \; , \\
  L_f \; , \;\;\;\; t\geq T \; .
 \end{array}\right.
\end{equation}

Finally, let us see how an unaccelerated 
rod looks from the point of view of an
accelerated observer. From (\ref{n3}), (\ref{tt1}) and (\ref{12.1}) we find 
\begin{equation}\label{17.2}
 L'(t')=\left\{
 \begin{array}{l}
  \displaystyle\frac{L_0}{\sqrt{1+{\rm sh}^2 at'/c}} \; , 
    \;\;\;\; 0\leq t'\leq T' \; , \\
      \\
  L_f \; , \;\;\;\; t'\geq T' \; . 
 \end{array}\right.
\end{equation}

The results (\ref{17}), (\ref{17.1}), (\ref{18}) and (\ref{17.2}) are
depicted in Fig.1. The parameters are chosen such that 
$a=9.81\; {\rm ms}^{-2}$, $aL_0/c^2 =0.5$, $\sqrt{1+(aT/c)^2}=4$, not with 
the intention to represent a realistic case, but rather to obtain results 
which will provide a synoptic view of the figure. The solid curves represent 
the lengths of the pushed and the pulled rods (\ref{17}) and
(\ref{17.1}), respectively. The 
short-dashed curve represents the approximative result (\ref{18}), while the 
long-dashed curve represents the length of an unaccelerated rod (\ref{17.2}).     

\figg{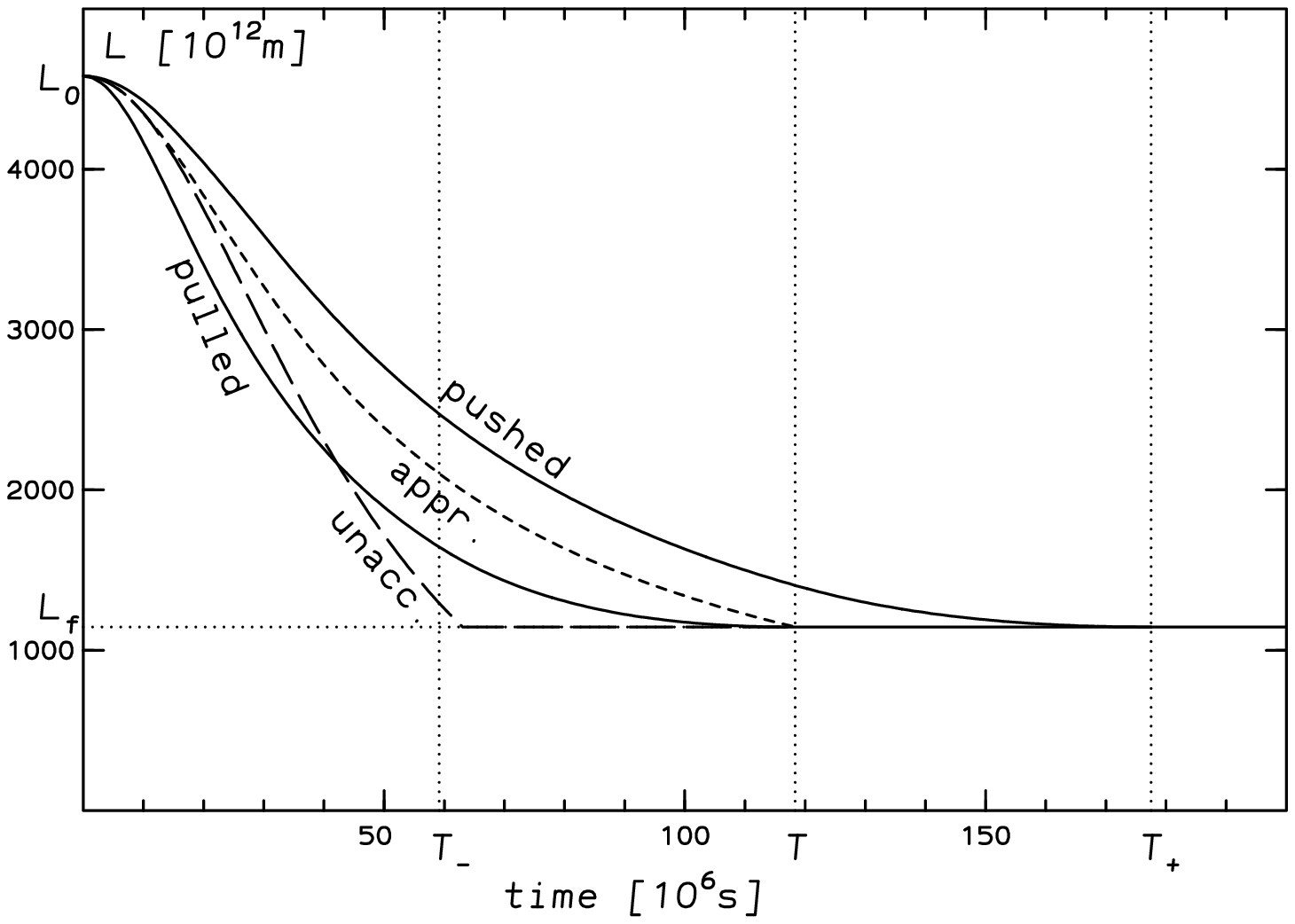}{7.7}{10.37}{The rod's length as a function of time.
The solid curves represent
the lengths of the pushed and the pulled rods
(\protect\ref{17}) and (\protect\ref{17.1}), respectively. The
short-dashed curve represent the approximative result (\protect\ref{18}), while the
long-dashed curve represent the length of an unaccelerated rod
(\protect\ref{17.2}).}{figr}

\section{Interpretation}

We see the inequivalence between an unaccelerated observer observing an 
accelerated rod and an accelerated observer observing an unaccelerated 
rod. For an 
accelerated observer, the acceleration lasts for a shorter time, as can be 
seen from (\ref{12.1}). The time dependence of the length of an
unaccelerated   
rod is given by (\ref{n3}), which can be understood as a simple 
generalization of the well-known Lorentz-Fitzgerald formula. On the 
other hand, a similar simple generalization (\ref{18})  
does not work for an accelerated rod. However, (\ref{18}) is a good approximation 
if $aL_0/c^2 \ll 1$. 

We also see that the results for a pulled rod are
meaningless if $aL_0/c^2 > 1$. This suggests that the rod cannot remain rigid 
under such conditions. To understand why is that so, we calculate the
velocity of the back end for a pulled rod. The result is 
\begin{equation}\label{21}
 \frac{d\tilde{x}_A (t)}{dt}=\frac{at}{\sqrt{(1-aL_0/c^2)^2 +(at/c)^2}} \; .
\end{equation}
We see that this velocity increases as $aL_0$ increases and reaches the 
velocity of light when $aL_0=c^2$. Since no part of the rod can exceed the 
velocity of light, the rod cannot remain rigid for $aL_0/c^2 > 1$. 
This is true for a pushed rod as well, as will be more clear from the 
discussion of Section 6. 

Although the time dependence of the rod's length depends on whether the rod
is pushed, pulled or unaccelerated, 
the final length and velocity after the forces 
are turned off do not depend on it. But what varies is the time needed to 
observe the termination of acceleration. An unaccelerated observer observes 
that the acceleration of the front end of the accelerated rod lasts longer 
than that of the back end. If the rod is pulled, it seems to him that the 
consequence (termination of acceleration of the back end) occurs before the 
cause (termination of acceleration of the front end). However, this is not 
a paradox, because there is a spacelike separation between these two
events, which is an artefact of the rigidity assumption. We make more 
comments on this in Section 6. 
An unaccelerated observer cannot actually know whether a rigid rod is 
pulled by an acceleration $a$, or is pushed by an acceleration $\tilde{a}$,
given by  
\begin{equation}\label{19}
 \tilde{a}=\frac{a}{1-aL_0/c^2} \; .
\end{equation}
This can be seen for example by replacing the acceleration in the first case
($0\leq t\leq T$) in 
(\ref{17}) by $\tilde{a}$ and comparing it with the first case 
($0\leq t\leq T_{-}$) in (\ref{17.1}). 
In particular, if the rod is pulled by acceleration $a=c^2/L_0$, for 
an unaccelerated  
observer it looks the same as it is pushed by acceleration $\tilde{a}=\infty$. 
If this pulling lasts time $T$, this is the same as the pushing lasts time 
$T_{-}=0$. 

Formula (\ref{19}) can be generalized to an arbitrary point on the rod. 
If acceleration $a$ is applied to the point $x_A'$, then this is
the same as acceleration $a(x')$ is applied to the point $x'$, where 
\begin{equation}\label{23}
 a(x')=\frac{a}{1+(x'-x_A')a/c^2} \; .
\end{equation}
The important consequence of this is that an observer in a uniformly
accelerated rocket does {\em not} feel a {\em homogeneous} 
inertial force, but rather an inertial  
force which decreases with $x'$, as given by (\ref{23}). This result 
is obtained also in [3].  
A long rigid rod 
is equivalent to a series of independent shorter rods, each having its 
own engine, but not with equal accelerations, but rather with accelerations 
which are related by formula (\ref{23}).  

Note also that one can replace $a$ by $a(x')$ in the first case 
($0\leq t_A \leq T$) in (\ref{12}) 
and thus obtain how the velocity of various points of a rod depends on time. 
The result coincides with the result obtained by Cavalleri and Spinelli [1].  
They found this result by solving a certain partial differential equation, so
our derivation is much simpler. In addition, our method allows a generalization 
to time-dependent accelerations as well.  

\section{A set of forces with various application points}

It is shown in [1] that if a time-independent 
force is applied to a {\em single} point on the rod, then this point moves 
in the same way as it would move if all mass
of the rod were concentrated in this point. However, if there are many forces     
directed along the length of a rigid rod, each applied to a different point 
on the
rod, then, obviously, all these points cannot move in the same way as they 
would move if all mass of the rod were concentrated in these points. We
remind the reader that it is enough only to find out how one particular
point of the rod moves, because the motion of the rest of rod is determined by 
the rigidity requirement. Thus the problem of many forces can be reduced 
to a problem of finding a point $x$ which moves in the same way as it would
move if all forces were applied to this point (in this section we 
omit a prime on $x'$, remembering that this is a coordinate on the rod in the 
accelerating rod's frame).    

Assume that $N$ forces $F_i$, $i=1,\ldots ,N$, are applied to the rod, each
applied to the point $x_i$. If all forces are of the same sign, then the rod 
(with a finite width) can be cut in $N$ pieces, each with a mass $m_i$ 
and each with only one applied force $F_i$, in 
such a way that the collection of pieces moves in the same way as the 
whole rod would move without the cutting. The masses of the pieces satisfy
\begin{equation}\label{g1}
 \sum_{i} m_i =m \; ,
\end{equation} 
where $m$ is the mass of the whole rod. We also introduce the notation 
\begin{equation}\label{g2}
 a_i =\frac{F_i}{m_i} \; , \;\;\;\; a=\frac{F}{m} \; ,  
\end{equation}
where $F=\sum_{i} F_i$. From (\ref{23}) it follows 
\begin{equation}\label{g3}
 \frac{c^2}{a_i}-\frac{c^2}{a_j}=x_i -x_j \; ,
\end{equation}
which leads to $N-1$ independent equations
\begin{equation}\label{g4}
 c^2\frac{m_{i+1}}{F_{i+1}} -c^2\frac{m_{i}}{F_{i}}=x_{i+1}-x_i \; , 
 \;\;\;\; i=1,\ldots ,N-1 \; .
\end{equation}
This, together with (\ref{g1}), makes a system of $N$ independent equations 
for $N$ unknown masses $m_i$, with the unique solution 
\begin{equation}\label{g5}
 m_i =\frac{F_i}{\sum_{j}F_j} \left[ m-\frac{1}{c^2}
 \sum_{k}F_k (x_k -x_i) \right] \; .
\end{equation}    
However, the masses $m_i$ are only auxiliary quantities. From (\ref{g5}) and 
the first equation in (\ref{g2}) we find 
\begin{equation}\label{g6}  
 a_i =\frac{\sum_{j}F_j}{m-\displaystyle\frac{1}{c^2}\sum_{k}F_k (x_k -x_i)} \; .  
\end{equation}
This is one of the final results, where masses $m_i$ do not appear. When 
$N=1$ or all forces are applied to the same point, (\ref{g6}) reduces to 
the already known result that the application point has the acceleration 
$a_i =F/m$. 

There is one point which accelerates in the same way as it 
would accelerate if all forces were applied to this point. This point 
$x$ is given by 
\begin{equation}\label{g7}
 \frac{c^2}{a}-\frac{c^2}{a_i}=x -x_i \; ,
\end{equation} 
so from (\ref{g7}), (\ref{g6}) and the second equation in (\ref{g2}) we find
\begin{equation}\label{g8}
 x=\frac{\sum_{k}F_k x_k}{\sum_{j}F_j} \; . 
\end{equation}

Formulas (\ref{g6}) and (\ref{g8}) are the main new results of this
section. We have derived them under the assumption that all forces $F_i$ are of
the same sign. However, we believe that 
this assumption is not crucial for the validity of 
(\ref{g6}) and (\ref{g8}). Since there must exist general formulas which 
reduce to (\ref{g6}) and (\ref{g8}) when all forces are of the same sign, 
we conjecture that these general formulas can be nothing else 
but (\ref{g6}) and
(\ref{g8}) themselves. For example, one could suspect that, in a general
formula,  
$F_k$ should be replaced by $|F_k |$, but one can 
discard such a possibility 
by considering the case when some forces of the different signs are applied to 
the same point. 

Are there any problems if all forces are not of the same sign? We assume that 
the square bracket in (\ref{g5}) is always positive (see the discussion 
connected with formula (\ref{21})). Thus, if some force $F_i$ is 
of the opposite sign with respect to the total force $F=\sum_{j}F_j$, then 
the corresponding mass $m_i$ is formally negative, which means that 
the rod cannot 
be cut in a way which was described at the beginning of this section. 
However, we believe 
that cutting the rod is not essential at all. 

From (\ref{g8}) one can also see that if all forces are not of the same
sign, then the point $x$ may not lie on the rod itself. This may look 
slightly peculiar, but is not inconsistent. 

Formulas (\ref{g6}) and (\ref{g8}) can be easily generalized to a 
continuous distribution of force. For example, (\ref{g8}) generalizes to 
\begin{equation}\label{g8.1}
 x=\frac{\int dy \: f(y)y}{\int dy \: f(y)} \; ,
\end{equation}
where $f(y)$ is a linear density of force, $f(y)=dF/dy$. 

\section{Discussion}

In this section we give a qualitative    
discussion of how the non-rigidity of realistic rods alters our
analysis and find conditions under which our analysis is
still valid, at least approximately. 

First, it is clear that, in general, the proper 
length of a uniformly accelerated rod will not be equal to the proper 
length of the same rod when it is not accelerated. For example, we expect that a 
pushed rod will be contracted, while a pulled rod will be elongated. This is 
not a relativistic effect, but rather a real change of a proper length.  
It is important, however, that if acceleration does not change with time, then 
this proper length does not change with time either. Therefore, all formulas of this 
article which describe a uniform acceleration during a long time interval,  
are correct if $L_0$ is understood as a proper length which depends on the 
acceleration and application point of the force, but not on the time of the 
accelerated frame. 

The dynamics of a rod when acceleration changes with time is more
complicated. 
However, some qualitative conclusions can be drawn without much effort. 
When acceleration is changed, the rod needs some relaxation time $\Delta t$ 
(here $t$ is time in the rod's accelerated frame) 
to reach a new equilibrium proper length which depends on the new 
acceleration. 
During this time we expect one or a few damped oscillations, 
so $\Delta t$ is of the order 
\begin{equation}\label{24} 
 \Delta t \approx L_0/v_s \; ,
\end{equation}
where $v_s$ is the velocity of propagation of a disturbance in a material. This 
velocity is equal to the velocity of sound (not of light) in a 
material of which is the rod made. If acceleration changes slowly enough, 
then we can use the adiabatic approximation, i.e., we can assume that 
the length of the rod is always equal to its equilibrium length  
which depends on the instantaneous acceleration. The small change of 
acceleration means that $\Delta a/a \ll 1$ during the relaxation time 
$\Delta t$, so from (\ref{24}) and the relation $\Delta a =\dot{a}\Delta t$,  
we find the criteria for the validity of the adiabatic approximation 
\begin{equation}\label{25}  
 \dot{a}\ll \frac{av_s}{L_0} \; .
\end{equation}
In practice, $\dot{a}$ is never infinite, i.e., the instantaneous changes of 
acceleration do not exist. 

As discussed in Section 4, owing to the rigidity assumption, it can happen that 
the consequence precedes the cause.  
In a more realistic calculation, which respects that $v_s$ is not larger
than $c$, this will not be the case. However, 
from the point of view of an unaccelerated inertial observer, it will still be true 
that the acceleration of a pushed rod lasts longer than that of a pulled one.

Let us now investigate the conditions under which a rod can 
be considered as approximately rigid, in the sense that the 
change of the proper length $\Delta L$ is much smaller than the proper
length $L_0$ of the unaccelerated rod. If the force is applied to the front 
end of the rod, then after the time given by (\ref{24})  
the back end will receive the information that it also 
has to accelerate. Therefore, the 
change of the proper length is of the order $\Delta L \approx a (\Delta t)^2 
\approx a L_0^2 /v_s^2$. Since $v_s < c$, the requirement $\Delta L /L_0
\ll 1$ leads to the requirement 
\begin{equation}\label{26}  
 a L_0 /c^2 \ll 1\; .
\end{equation}
In that case, (\ref{18}) is a good approximation and the difference 
between the pushed and the pulled rod is negligible.  

Finally, note that the results of Sections 2 and 3 are exact if labels $A$ and $B$ 
do not refer to fixed points on the rod, but one of the labels refers to the
observer whose acceleration {\em is} given by $a(t')$ and the other refers
to the point the distance of which from the first point {\em is} $L_0$, 
as seen by the observer at the first point.  


\section{Conclusion}

In this article we have relativistically solved a general problem of motion of 
an arbitrary point of a rigid rod accelerated by a time-dependent 
force applied to a single point, for the case 
when the force and velocity are directed along the rod's length. The time-dependence 
of a rod's relativistic length depends on the application point 
of the force, but the final velocity and length after the termination of 
acceleration do not depend on it. An observer on a uniformly accelerated rod 
does not feel a homogeneous inertial force, but rather an inertial force
which decreases in the direction of acceleration.  
Formula (\ref{g8}) determines the motion of a rigid rod when 
many time-independent forces directed along the rod's length 
are applied to various points. The case of many time-dependent 
forces applied to various points is more complicated, so 
we have not considered this case. In addition, we have given a qualitative 
discussion of how the non-rigidity of realistic rods alters our 
analysis and found conditions under which our analysis is 
still valid, at least approximately.

\section*{Acknowledgement}

The author is grateful to Damir Stoi\'{c} for some extremely useful discussions.
This work was supported by the Ministry of Science and Technology of the 
Republic of Croatia under Contract No. 00980102.


\begin{thebibliography}{99}
\bibitem{caval}
 G. Cavalleri and G. Spinelli, ``Does a Rod, Pushed by a Force, 
 Accelerate Less than the Same Rod Pulled by the Same Force?,"
 Nuovo Cimento B {\bf 66} (1), 11-20 (1970).
\bibitem{nord}
 K. Nordtvedt, ``The equivalence principle and the question
 of weight," Am. J. Phys. {\bf 43} (3), 256-257 (1975).
\bibitem{gron}
 \mbox{\rlap{\,/}O.} Gr\rlap/on, ``Acceleration and weight 
 of extended bodies in the theory of relativity," 
 Am. J. Phys. {\bf 45} (1), 65-70 (1977).    
\bibitem{nelson}
 R. A. Nelson, ``Generalized Lorentz transformation for an accelerated,
 rotating frame of reference," 
 J. Math. Phys. {\bf 28} (10), 2379-2383 (1987).
\end{thebibliography}
\end{document}